\documentclass[letter]{aa}

\usepackage{bm} % bold math

\usepackage[T1]{fontenc}
\usepackage[utf8]{inputenc}

\usepackage{ae,aecompl}
\usepackage{graphicx}   % Including figure files
\usepackage{amsmath}    % Advanced maths commands
\usepackage{amssymb}    % Extra maths symbols

\usepackage{mathtools}
\usepackage{natbib}
\bibpunct{(}{)}{;}{a}{}{,} % to follow the A&A style

\usepackage[breaklinks=true]{hyperref}

\hypersetup{
   colorlinks=true,
   linkcolor=blue,
   citecolor=blue,
   urlcolor=blue,
}

\usepackage{enumerate}

\usepackage{array}
\newcolumntype{L}[1]{>{\raggedright\let\newline\\
\arraybackslash\hspace{0pt}}m{#1}}
\newcolumntype{C}[1]{>{\centering\let\newline\\
\arraybackslash\hspace{0pt}}m{#1}}
\newcolumntype{R}[1]{>{\raggedleft\let\newline\\
\arraybackslash\hspace{0pt}}m{#1}}

\def\Msun{\mbox{M$_\odot$}}

\def\mst{\mbox{$M_{\star}$}}

\def\lsim{\mathrel{\rlap{\lower3.5pt\hbox{\hskip0.5pt$\sim$}}
    \raise0.5pt\hbox{$<$}}}                % less than or approx. symbol
\def\gsim{~\rlap{$>$}{\lower 1.0ex\hbox{$\sim$}}}

\def\SN{\mbox{$S/N$}}

\def\Fig{\mbox{Figure~}}

\def\UCMG{\mbox{\textsc{UCMG}}}
\def\UCMGs{\mbox{\textsc{UCMGs}}}

\def\NSMGs{\mbox{\textsc{NSMGs}}}

\def\Re{\mbox{$R_{\rm e}$}}
\def\Te{\mbox{$\Theta_{\rm e}$}}

\def\sqd{\mbox{~sq.~deg.}}
\def\Rvir{\mbox{\rm $R_{\rm vir}$}}

\def\Pcl{\mbox{$P_{\rm cl}$}}

\def\perc{\mbox{\%}}

\defcitealias{Tortora+18_UCMGs}{T18}
\defcitealias{Scognamiglio+20_UCMGs}{S20}

\begin{document}

\title{Nature versus nurture: relic nature and environment of the most massive passive galaxies at $z < 0.5$}

\author{C.~Tortora\inst{\ref{inst1}}
    \and
 N.~R.~Napolitano\inst{\ref{inst2},\ref{inst3}}
 \and
 M.~Radovich\inst{\ref{inst4}}
  \and
 C.~Spiniello\inst{\ref{inst3},\ref{inst5}}
  \and
  L.~Hunt\inst{\ref{inst1}}
  \and
  N.~Roy\inst{\ref{inst2}}
  \and
  L.~Moscardini\inst{\ref{inst6},\ref{inst7},\ref{inst8}}
  \and
  D.~Scognamiglio\inst{\ref{inst9}}
  \and
  M.~Spavone\inst{\ref{inst3}}
  \and
  M.~Brescia\inst{\ref{inst3}}
  \and
  S.~Cavuoti\inst{\ref{inst3},\ref{inst10}}
  \and
  G.~D`Ago\inst{\ref{inst11}}
  \and
  G.~Longo\inst{\ref{inst10}}
  \and
  F.~Bellagamba\inst{\ref{inst6},\ref{inst7}}
  \and
  M.~Maturi\inst{\ref{inst12},\ref{inst13}}
  \and
  M.~Roncarelli\inst{\ref{inst6},\ref{inst7}}}

%\email{crescenzo.tortora@inaf.it}

\institute{ INAF -- Osservatorio Astrofisico di Arcetri, Largo
Enrico Fermi 5, 50125, Firenze, Italy,
\email{crescenzo.tortora@inaf.it}\label{inst1} \and School of
Physics and Astronomy, Sun Yat-sen University Zhuhai Campus, Daxue
Road 2, 519082 - Tangjia, Zhuhai, Guangdong, P.R.
China\label{inst2} \and INAF -- Osservatorio Astronomico di
Capodimonte, Salita Moiariello 16, 80131 - Napoli,
Italy\label{inst3}  \and INAF -- Osservatorio Astronomico di
Padova, Vicolo Osservatorio 5, 35122 - Padova, Italy\label{inst4}
\and Department of Physics, University of Oxford, Keble Road,
Oxford OX1 3RH, UK\label{inst5} \and Dipartimento di Fisica e
Astronomia -- Alma Mater Studiorum Universit\`{a} di Bologna, via
Piero Gobetti 93/2, I-40129 Bologna, Italy\label{inst6}  \and INAF
-- Osservatorio di Astrofisica e Scienza dello Spazio di Bologna,
via Piero Gobetti 93/3, I-40129 Bologna, Italy\label{inst7} \and
INFN -- Sezione di Bologna, viale Berti-Pichat 6/2, I-40127
Bologna, Italy\label{inst8} \and Argelander-Institut f\"ur
Astronomie, Auf dem H\"ugel 71, D-53121 - Bonn,
Germany\label{inst9} \and Dipartimento di Scienze Fisiche,
Universit\`{a} di Napoli Federico II, Compl. Univ. Monte S.
Angelo, 80126 - Napoli, Italy\label{inst10} \and Instituto de
Astrof\'isica Pontificia Universidad Cat\'olica de Chile, Avenida
Vicu\~na Mackenna, 4860 - Santiago, Chile\label{inst11} \and
Zentrum f\"ur Astronomie, Universitat\"at Heidelberg,
Philosophenweg 12, D-69120 Heidelberg, Germany\label{inst12} \and
Institut f\"ur Theoretische Physik, Ruprecht-Karls-Universit\"at
Heidelberg, Philosophenweg 16, D--69120 Heidelberg,
Germany\label{inst13}}

\date{Received XXX; accepted YYY}

\abstract{Relic galaxies are thought to be the progenitors of
high-redshift red nuggets that for some reason missed the channels of size
growth and evolved passively and undisturbed since the first star
formation burst (at $z>2$). These local ultracompact old galaxies
are unique laboratories for studying the star formation processes at
high redshift and thus the early stage of galaxy formation
scenarios. Counterintuitively, theoretical and observational
studies indicate that relics are more common in denser
environments, where merging events predominate. To verify this scenario,
we compared the number counts of a sample of ultracompact
massive galaxies (\UCMGs) selected within the third data release
of the Kilo Degree Survey, that is, systems with sizes $\Re < 1.5 \,
\rm kpc$ and stellar masses $\mst > 8 \times 10^{10}\, \rm \Msun$,
with the number counts of galaxies with the same masses but normal sizes in
field and cluster environments. Based on their optical and
near-infrared colors, these \UCMGs\ are likely to be mainly old,
and hence representative of the relic population. We find that
both \UCMGs\ and normal-size galaxies are more abundant in
clusters and their relative fraction depends only mildly on the
global environment, with denser environments penalizing the
survival of relics. Hence, \UCMGs\ (and likely relics overall) are
not special because of the environment effect on their nurture,
but rather they are just a product of the stochasticity of the
merging processes regardless of the global environment in which
they live. }


\titlerunning{Nature versus nurture in \UCMGs}
\authorrunning{C.~Tortora}

\maketitle

\section{Introduction}\label{sec:intro}

According to the current understanding of galaxy formation, cosmic
structures are formed in a hierarchical fashion. At the very early
stages of the universe history, gas condenses within the
primordial dark matter perturbations
(\citealt{Blumenthal+84_nature}; \citealt{Springel+05_Nature}),
forming the first stars and thus the first galaxies. With time,
these objects grow through mergers, generating increasingly larger
galaxies, and experience physical transformations influenced by
the wide range of environments that they live in
(\citealt{Daddi+05}; \citealt{Trujillo+07}). Determining the
role of internal processes and environment (this last including
mergers), which astronomers refer to as the ``nature versus
nurture'' problem (e.g., \citealt{Irwin95}), is crucial for
understanding the formation and evolution of galaxies.

Recent studies seem to suggest that the most massive and passive
galaxies with $\mst \gsim \, 5 \times 10^{10}\, \rm \Msun$ are
formed through a {\it \textup{two-phase formation scenario}}
(\citealt{Oser+10}). During the first phase, an intense and quick
burst of star formation at $z > 2$ forms the bulk of the central mass and after the star formation quenches,
generates compact massive quiescent galaxies (red nuggets). Then, during
a more extended second phase, mergers and gas inflows cause a
dramatic size growth despite a very slight change in mass (e.g.,
\citealt{Hilz+13}; \citealt{Tortora+14_DMevol}). Red nuggets, $ \text{which are about four times}$ smaller than local massive galaxies, are very
common at $z > 2$, but are expected to be rare in the local universe
\citep{vanDokkum+06,Cimatti+08,Bezanson+09}. Because of the
stochastic nature of mergers, simulations predict that only $\sim
1-10\%$ of the red nuggets evolve undisturbed until the present cosmic epoch,
with a very slight change in size, and therefore represent outliers of
the local size-mass relation
\citep{Hopkins+09_DELGN_IV,Quilis_Trujillo13}; these peculiar
galaxies are named ``relics''. The number of relics depends on the
physical processes acting during the second phase, and in
particular, on the relative contribution of major and minor galaxy
mergers. Therefore, comparing relic abundance and properties with
those of normal-size equally massive galaxies allows us to
measure the merging effect. The stars that formed in red nuggets are
thought to be the in situ populations living in the core of
local giant ellipticals. However, this population is mixed with
the accreted population that formed during mergers and inflows. With relics, which lack the accreted component, we can
therefore probe the processes that shape the galaxy formation at high
redshift with a precision that is only attainable for the nearby universe,
in order to separate nature from nurture.

In the past few years, several different studies found and
characterized compact massive galaxies and relics in local
environments. The only difference between the former and the
latter is the age of the stellar population, which for relics is
as old as the universe. They extended the analysis up to $z =
0.7$, where the number counts are expected to be higher than in
the local universe (e.g., \citealt{Tortora+18_UCMGs} versus
\citealt{Ferre-Mateu+17}) and high enough to study the evolution
of these galaxies within different environments.

While the number of discovered compact galaxies is increasing at
$z \lsim 0.5$ \citep[e.g.,][and references
therein]{Tortora+18_UCMGs}, there is still an open debate about possible selection biases related to the environment. The results of large sky surveys such as the Sloan Digital Sky
Survey (SDSS\footnote{https://www.sdss.org/}) show a sharp decline
in compact galaxy number density of more than three orders of
magnitude below the high-redshift values \citep[$z \sim
2$;][]{Trujillo+09_superdense, Taylor+10_compacts}. In contrast,
data in nearby clusters indicate a number density of two orders of
magnitude above the SDSS one, which is comparable with the number density
at high redshift \citep{Valentinuzzi+10_WINGS,Poggianti+13_low_z,
Poggianti+13_evol}.

Thus, the following questions need answers. Is cluster environment
favoring the formation of relics? Are these rare systems
descendants of red nuggets or formed preferentially in clusters
from larger galaxies deprived of their outer stellar populations?

Using 271 compact, massive ($\mst \gsim 10^{10}\, \Msun$) and
quiescent galaxies at $0.1 < z \lsim 0.6$ in the COSMOS field,
\cite{Damjanov+15_env_compacts} have demonstrated that compact
quiescent galaxies populate a similar range of environment as the
parent population of equally massive quiescent galaxies. The
numbers of these compact and normal-size systems are not yet
statistically significant, however, for an investigation of the
environmental dependence of the most massive, red, and extremely
compact systems, that are the most likely direct descendants of
high-z red nuggets.

In this letter we make a crucial step forward in understanding the
role of the environment in relic formation and evolution. As realistic markers of the relic population we use the largest
available sample ($\sim1000$ objects) of ultracompact massive
galaxies (\UCMGs\ hereafter), which are defined as the most compact ($\Re <
1.5\, \rm kpc$) and most massive ($\mst > 8 \times 10^{10}\,
\Msun$) red galaxies at $z < 0.5$. These \UCMGs\ have been found
by \citet[hereafter T18]{Tortora+18_UCMGs} within the footprint of
the VLT Survey Telescope (VST) Kilo Degree Survey (KiDS, \citealt{deJong+17_KiDS_DR3}).
We investigate their number counts in terms of the environment
(fields versus clusters) and compare our findings with a KiDS parent
sample of normal-size massive galaxies. We adopt a cosmological
model with $(\Omega_{m},\Omega_{\Lambda},h)=(0.3,0.7,0.7)$, where
$h = H_{0}/100 \, \textrm{km} \, \textrm{s}^{-1} \,
\textrm{Mpc}^{-1}$ (\citealt{Komatsu+11_WMAP7}).

\section{Galaxy samples}\label{sec:data_and_selection}

The galaxy selection started from the KiDS multiband source
catalog that is included in the third KiDS Data Release
(KiDS--DR3, \citealt{deJong+17_KiDS_DR3}). After masking of bad
areas, we collected a catalog of $\text{about } 5$ million
galaxies within an effective area of 333 \sqd. The photometric
catalog includes $u$-, $g$-, $r$-, and $i$-band magnitudes
(\citealt{deJong+17_KiDS_DR3}), structural parameters obtained
from the point spread function (PSF)-convolved fit of the S\'ersic
profile (\citealt{Roy+18}), photometric redshifts
(\citealt{Cavuoti+17_KiDS}), and stellar masses derived from
spectral energy distribution (SED) fitting of single-burst
(\citealt{BC03}) stellar population synthesis theoretical models
(\citetalias{Tortora+18_UCMGs}). We complemented these data with a
galaxy classification based on SED fitting and VIKING $J$ and $K$
magnitudes, with which we set up a color cut in the $g$-$J$ versus
$J$-$K$ plane to further remove stellar contaminants and very blue
galaxies, as discussed in \citetalias{Tortora+18_UCMGs}. Because
most of our (compact and more extended) galaxies are elliptical,
we expect that the number of galaxies that are misclassified as
stars from the KiDS pipeline is very small. For more details about
data extraction, quality checks, and sample selection, we refer to
our previous papers (\citealt{Roy+18};
\citealt{Tortora+16_compacts_KiDS}; \citetalias{Tortora+18_UCMGs};
\citealt[S20 hereafter]{Scognamiglio+20_UCMGs}).

Here, we only use galaxies with good surface photometry fits,
selecting a cumulative $r$-band signal-to-noise ratio, $\SN
> 50$,  a good $\chi^2$ ($< 1.5$), and realistic structural parameters in
g, r, and i bands (S\'ersic index $n
> 0.5$, axis ratios $q
> 0.1,$ and effective radius $\Te > 0.05''$); these criteria also reduce the
contaminations by misclassified stars, disk-on galaxies, and
systems with spiral arms.

We selected a sample of 104383 massive galaxies with $\mst
> 8 \times 10^{10}\, \rm \Msun$ at redshifts $z < 0.5$. According
to their median effective radius \Re, calculated as the median of
the $g$-, $r$-, and $i$-band \Re, we then classified them into two
separate samples: normal-size massive galaxies (\NSMGs), and
ultracompact massive galaxies (\UCMGs):
\begin{itemize}
\item {\it \NSMGs:} 103388 objects with a median effective radius $\Re \geq 1.5\, \rm kpc$.
\item {\it \UCMGs:} 995 objects with a median effective radius $\Re < 1.5\, \rm kpc$ instead.
\end{itemize}

The search for cluster candidates was made using the algorithm
called Adaptive Matched Identifier of Clustered Objects
(\texttt{AMICO}) (\citealt{Bellagamba+18_AMICO}), which applies an
optimal filter to select galaxy overdensities in a catalog with
coordinates, photometric redshifts, and magnitudes of galaxies. We
applied this algorithm to KiDS--DR3
(\citealt{Bellagamba+19_AMICO_KiDS-DR3};
\citealt{Maturi+19_AMICO}; Radovich et al., submitted). Each
galaxy is tagged with its distance from the cluster center and a
membership probability, \Pcl, that is, the probability (from 0 to
1) to be a cluster member. An estimate for the cluster virial
radius (\Rvir) is also available. In the following, we limit our
analysis to galaxies with $\Pcl > 0.2$.

\section{Galaxy number counts and environment}\label{sec:results}

\begin{figure*}
\centering
\includegraphics[width=0.45\textwidth]{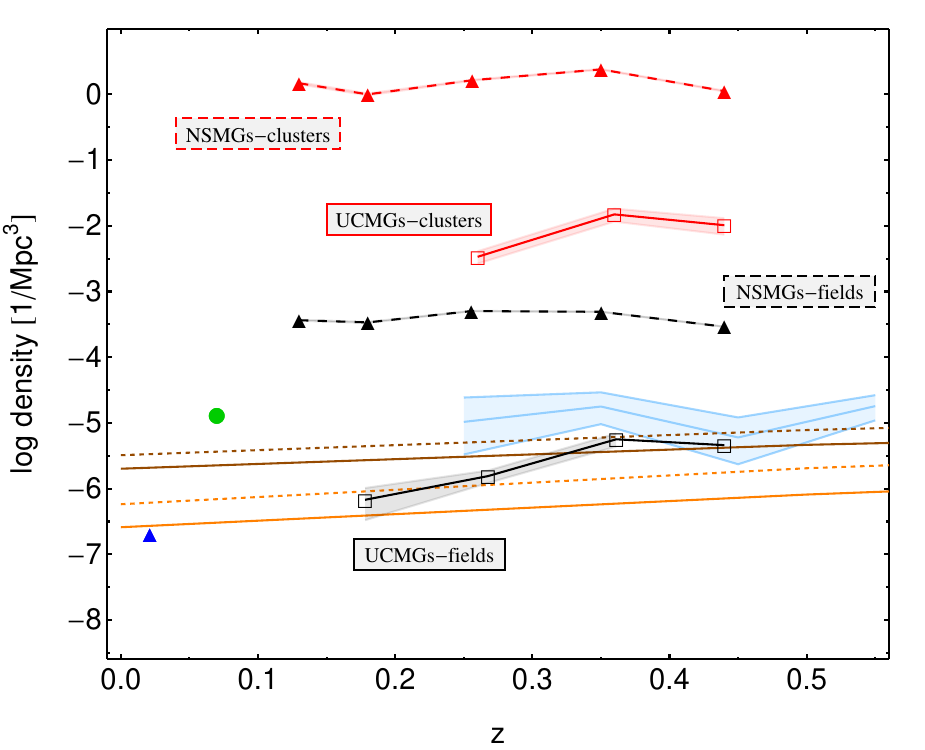}
\includegraphics[width=0.45\textwidth]{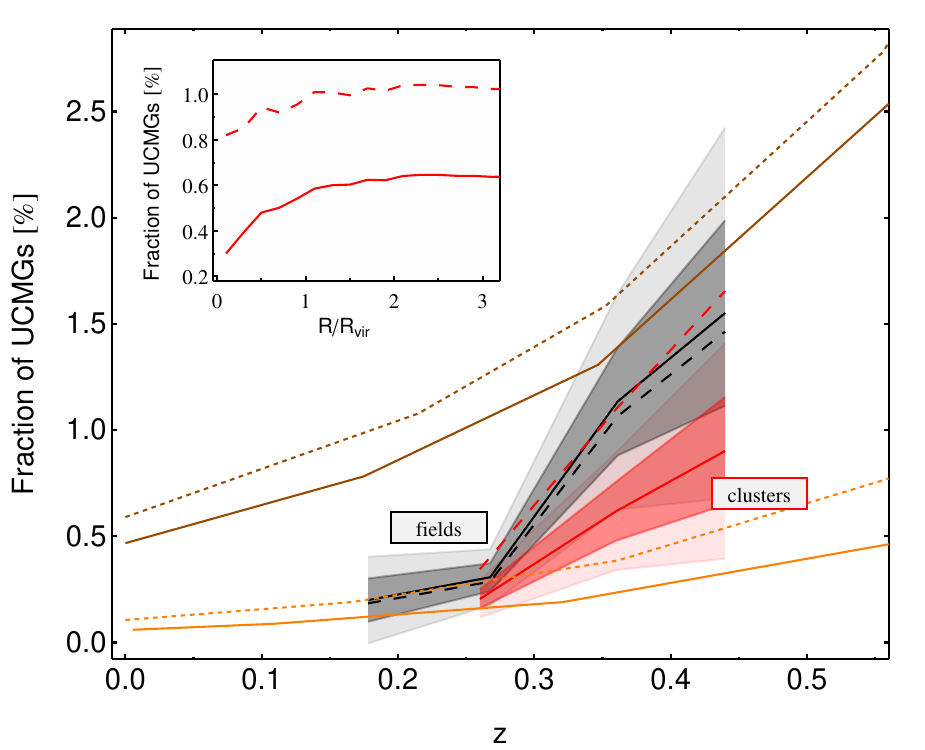}
\caption{{\it Left panel.} Number density of \UCMGs\ (solid line,
squared points) and \NSMGs\ (dashed lines, triangles) as a
function of redshift and environment (black lines for field and
red for cluster galaxies). Shaded regions correspond to 1$\sigma$
errors, accounting for Poisson noise, cosmic variance, and
uncertainties in size and mass selection, but neglecting the
nominal uncertainties on the photometric redshifts
(\citetalias{Scognamiglio+20_UCMGs}). The cyan line refers to
number densities for \UCMGs\ in the COSMOS area by
\citet{Damjanov+15_compacts}. The green point and blue triangle
are the results for compact galaxies from
\citet{Poggianti+13_low_z} and \citet{Ferre-Mateu+17},
respectively. Orange and brown curves are extracted from
\citet{Quilis_Trujillo13}: dashed and solid lines refer to
\citet{Guo+11_sims} and \citet{Guo+13_sims} SAMs, respectively,
while orange (brown) lines are for galaxies that have increased their
mass by less than 10 (30)\perc. No selection in environment is
performed in such simulations. {\it Right panel.} Fraction of
\UCMGs, calculated with respect to the total parent population  in
fields (black) and clusters (red), as a function of redshift. Dark
(light) shaded regions show 1$\sigma$ (2$\sigma$) errors in each
redshift bin. Dashed black and red lines are for \UCMGs\ with
$\log \mst/\Msun \leq 11.2$ dex in field and clusters,
respectively. The \citet{Guo+11_sims} and \citet{Guo+13_sims}
results are also plotted. {\it Inset.} We plot the fraction of
\UCMGs\ in clusters as a function of the distance R from the
center in units of \Rvir. The dashed line is for \UCMGs\ with $\log
\mst/\Msun \leq 11.2$ dex.} \label{fig:densities}
\end{figure*}

In this section we discuss the numerical abundance of \UCMGs\ and
\NSMGs, the fraction of \UCMGs,\, and their distribution as a
function of the environment. We then compare our results with
literature and finally extend them to relic galaxies.

%\LEt{a single sentence does not constitute a paragraph. Please add
%at least one other sentence or remove}.

\subsection{Number density calculation}

Following \citetalias{Tortora+18_UCMGs}, we started by determining
number densities for \NSMGs\ and \UCMGs\ within the KiDS effective
area, regardless of the environment. We binned galaxies according
to redshift. As in \citetalias{Tortora+18_UCMGs}, we optimized the
redshift bins for the \UCMG\ sample, setting a width of $0.1$,
except for the lowest-z bin that corresponds to the redshift
interval $(0.15-0.2)$; for \NSMGs\ we also added the interval
$(0-0.15)$. We multiplied the number of candidates by $f_{\rm
area} = A_{\rm sky}/A_{\rm survey}$, where $A_{\rm sky}$
($=41253$\sqd) is the full-sky area and $A_{\rm survey}$ ($=
333$\sqd) is the effective KiDS area. The density was derived by
dividing by the all-sky comoving volume corresponding to each
redshift bin.

To obtain galaxy counts in clusters, we first selected cluster
galaxies in each redshift bin, with $\Pcl > 0.2$, within 1 $R_{\rm
vir}$, and we weighted them according to \Pcl, giving more weight to galaxies that are more likely cluster members. For
both \NSMGs\ and \UCMGs,\ the total number of galaxies in each
redshift bin was then divided by the sum of the comoving volumes within
\Rvir\ of all the clusters and in that redshift bin.

Finally, we obtained the number of \NSMGs\ or \UCMGs\ in the
fields by subtracting the cluster members from the total number of
\NSMGs\ or \UCMGs. Similarly, the comoving volume was determined
by subtracting the comoving volume occupied by clusters from the
total volume. Clusters occupy a volume a factor $\sim 4 \times
10^{-5}$ smaller than the whole effective volume we analyzed.

\subsection{Number density and environment}

In the left panel of \Fig\ref{fig:densities} the number densities
for field and cluster \UCMGs\ are compared with those of \NSMGs\
in the same environments. Number counts for field \UCMGs\ are
found to decrease with cosmic time from $\sim 9 \times 10^{-6} \,
\rm Mpc^{-3}$ at $z \sim 0.5$ to $\sim 10^{-6} \, \rm Mpc^{-3}$ at
$z \sim 0.15$. This corresponds to a decrease of $ \text{about
nine}$ times in about 3 Gyr (\citetalias{Tortora+18_UCMGs};
\citetalias{Scognamiglio+20_UCMGs}). We found that \UCMGs\ within
1 \Rvir\ of the clusters are more abundant than field \UCMGs\ with
numbers counts of $\sim 5.7 \times 10^{-3} \, \rm Mpc^{-3}$ at $z
\gsim 0.25$. The trend with redshift seems to be similar to the
trend in the field, but we did not find any \UCMGs\ in clusters
below $z=0.25$.

In the same plot, we also show the number counts for \NSMGs\ in
fields and clusters. The number counts are systematically larger
than those of \UCMGs\ in the same environment. The number counts
of \NSMGs\ are constant with redshift in clusters ($\sim 1.5 \,
\rm Mpc^{-3}$) and in the field ($\sim 4 \times 10^{-4} \, \rm
Mpc^{-3}$). This suggests that the fraction of recently formed
massive red galaxies is negligible. This is consistently with
previous results (e.g., \citealt{Cassata+13}).

%more abundant

We also compared the results for \UCMGs\ with independent
findings. The cyan region in the left panel of
\Fig\ref{fig:densities} shows number densities of galaxies in the
COSMOS survey \citep{Damjanov+15_compacts}. Our results for field
\UCMGs\ (or equivalently, those determined from the whole survey
area) are consistent with COSMOS number counts in the highest
redshift bin, but are systematically lower at lower z, with
differences of about one order of magnitude in the lowest-z bin.
Below $z \sim 0.2,$ our number counts decrease by one order of
magnitude and appear to follow the direction of the local estimate
from \cite{Ferre-Mateu+17}, who found a number density of $\sim 2
\times 10^{-7}\, \rm Mpc^{-3}$ within a sphere of radius 106 Mpc
from us. Instead, over an area of $38\sqd$, biased toward dense
cluster environments, \cite{Poggianti+13_low_z} have found four
galaxies (older than $8\, \rm Gyr$) that fulfill our criteria,
corresponding to a large number density of $\sim 10^{-5}\, \rm
Mpc^{-3}$. KiDS number densities in cluster environments are three
orders of magnitude larger than their values. The source of this
large discrepancy is related to a different strategy used to
normalize the number counts (our cluster volumes versus their
volume calculated within the area of $38\sqd$). We finally
compared our results with the density of compact galaxies
extracted from semi-analytical models (SAMs) based on Millennium
N-body simulations (\citealt{Guo+11_sims, Guo+13_sims}). There is
a clear overlap with number density of field \UCMGs.

\subsection{Fraction of \UCMGs\ and environment}

We report here the total absolute numbers (weighted according to
\Pcl) of \NSMGs\ and \UCMGs\ in clusters and their fraction with
respect to the their total galaxy population (including field and
cluster systems). The number of \NSMGs\ in clusters is $\sim
22246$, which corresponds to $\sim 22\perc$ of the total number of
\NSMGs. Instead, the number of cluster \UCMGs\ is $\sim 135$, which is
$\sim 14\perc$ of the total number of \UCMGs. This trend is made
more robust when we separate \NSMGs\ into four \Re\ bins $(1.5-3)$,
$(3-5)$, $(5-7),$ and $(7-50)$ kpc. This shows that galaxies in
clusters are 18, 19, 19, and 24\perc\ of the total in these \Re\
bin, respectively. The tendency for higher
mass galaxies to be preferentially found in clusters is therefore clear. The tendency is expected when we consider that higher density regions favor mergers, and thus the formation of larger galaxies.

The average size of the galaxies is only slightly larger in
clusters ($\sim 5\perc$ more). It reaches an increment of $\sim
11\perc$ at $\mst > 2 \times 10^{11}\, \rm \Msun$, in agreement
with the negligible or mild dependence found at low and
intermediate redshift (\citealt{Huertas-Company+13_local,
Huertas-Company+13_evol}; \citealt{Lani+13_env_size}).

In the right panel of \Fig\ref{fig:densities} we plot the
fraction of \UCMGs\ with respect to the total galaxy population
(including both \NSMGs\ and \UCMGs) in fields and clusters as a function of redshift. The left panel of
the same figure shows that the fraction of \UCMGs\ decreases in the last 3
Gyr, which is expected when we consider that the probabilitly of merging
increases with time. The fraction of compact systems in clusters
is smaller than that of the same galaxies in the field, and
it is consistent within the typical uncertainties. As mergers are more
likely to occur in clusters, the fraction of \UCMGs\ that merge
to form \NSMGs\ is larger. These results clearly show that \UCMGs\
are more abundant in clusters because the parent sample of massive
galaxies is more abundant there, and not because of their compactness. Mirroring
the comparison made in terms of number counts, the two sets of
simulations in \cite{Quilis_Trujillo13} present a shallower
evolution with redshift and bracket our results.

In the inset of the right panel of \Fig\ref{fig:densities}, we
also plot the fraction of \UCMGs\ in clusters, calculated within a
distance $\rm R$ from the cluster center, given in units of \Rvir.
The fraction is very low in the very central regions of clusters
(i.e., $\sim 0.3\perc$) and when galaxies at larger
distances from the center are included, it increases to $\sim 0.6\perc$ when all the
galaxies within $\sim 3 \Rvir$ are considered. This means that not
only are \UCMGs\ less common in clusters than in
fields, but their fraction is also halved in the central regions when
compared with more peripheral regions.

Although the ratio between the average stellar masses of \UCMGs\
and \NSMGs\ remains almost constant with redshift, 946 out of 995
\UCMGs\ ($\sim 95\perc$) have $\log \mst/\Msun \leq 11.2$ dex. We
therefore also calculated galaxy fractions using only \UCMGs\ and
\NSMGs\ in this mass range. In this case, \UCMGs\ in clusters are
slightly more abundant. Their mean fractions coincide with those
for field \UCMGs\ and vary from $\sim 1$ to $\sim 0.8\perc$ from
the peripheries to the cluster centers. Nevertheless, our main
conclusions are entirely unaffected.

\subsection{Color dependence and relic candidates}

Finally, in this section we evaluate the effect of color on our
galaxy selections. The majority of galaxies in our samples have
red optical colors, which resemble spectral templates of
ellipticals (\citealt{Ilbert+06}). These red galaxies represent
93\perc\ of \NSMGs\ (96\perc\ in clusters) and 98\perc\ of \UCMGs\
(98\perc\ in clusters). This is expected because the galaxy
population at high mass is dominated by passive and red systems
(\citealt{Kauffmann+03}; \citealt{PengYJ+10};
\citealt{Vulcani+15}). When only these red galaxies are included,
the number densities slightly decrease, and the ratio between
\UCMGs\ and the whole galaxy populations is left unchanged.
Inspecting more restrictive color cuts, we still find a negligible
effect on our results.

A limit to the reddest galaxy population means a further
reduction of the contamination by young systems. The negligible
effect of color selection on the fraction of \UCMGs\ can extend our results to the reddest and oldest \UCMGs, that is, to the
relic galaxies. Therefore, relic galaxies populate a distribution
of environments similar to their parent massive and passive galaxy
population.

\section{Discussion and conclusions}\label{sec:conclusions}

We have collected the largest sample of \UCMGs\ at $z < 0.5$
within the 333\sqd\ of the third data release of the KiDS survey
and investigated their abundance as a function of the environment.
The environment was characterized by selecting galaxies in clusters
and in the field, and the abundances were compared with those of their
parent population of massive galaxies with larger sizes (\NSMGs).

We showed that \NSMGs\ and \UCMGs\ populate a similar range of
environments: they are both more abundant in clusters, but their
ratio is almost independent of the environment. In more detail,
\UCMGs\ are mildly less abundant in clusters, and their fraction
with respect to the total massive galaxy population is halved in
the very central part of clusters (from $\sim 0.6$ to $0.3\perc$).
We also showed that the results do not depend on galaxy colors,
based on which we extended these findings to the most likely
candidates to be relic galaxies (the reddest and oldest \UCMGs).
{\it \textup{This result refutes the misconception that relic
galaxies are more abundant in denser environments than relic
galaxies located in the field.}} Relic galaxies are more abundant
in denser environments because they are part of the massive and
passive galaxy population, which is preferentially located in
clusters. Our analysis focused on the most massive and compact
galaxies and complements similar findings obtained by
\cite{Damjanov+15_env_compacts} that were based instead on a
smaller number of galaxies. These authors adopted more relaxed
criteria on stellar mass ($\mst
> 10^{10}\, \rm \Msun$ versus the actual $\mst > 8 \times 10^{10}\,
\rm \Msun$) and size for \UCMGs\ (the \citealt{Barro+13} selection
criterion versus the $\Re < 1.5 \, \rm kpc$ criterion adopted
here). Our results appear to disagree with those of
\cite{PeraltadeArriba+16}, who reported that $z \sim 0$ relics
with $\mst \gsim 10^{10}\, \rm \Msun$ prefer denser environments;
however, the different mass range considered might drive this
discrepancy.

The implications of these results are very relevant for the
two-phase formation scenario. First, the smaller fraction of relic
candidates found in the clusters at $z \lsim 0.5$  and in
particular in the cluster cores disfavors the hypothesis that they
formed autochthonously, that is, the possibility that they are
formed by environmental processes that have compacted preexisting
larger cluster galaxies. Instead, the probability of being
involved in a merger is higher in these dense environments, which
penalizes the survival of relics. Second, when such environmental
physical processes are excluded, the rarity of relic galaxies can
only be explained by the stochastic nature of mergers in any type
of environment. Minor mergers drive the size evolution of the most
massive and largest galaxies in the local universe
(\citealt{Trujillo+07}; \citealt{Hilz+13};
\citealt{Tortora+18_KiDS_DMevol}), but because of stochasticity,
they miss the few relic galaxies (\citealt{Oser+10};
\citealt{Martin-Navarro+15_IMF_relic,
Martin-Navarro+15_IMF_variation}; \citealt{Ferre-Mateu+17}).

With the ongoing INSPIRE Project (Spiniello et al. 2020, to be
submitted), we will investigate the stellar populations,
structural properties, and environment dependence of a smaller but
purer sample of spectroscopically validated relic galaxies. This will add
new pieces of information with which the two-phase
formation scenario can be tested unambiguously.

%%%%%%%%%%%%%%%%%%%%%%%%%%%%%%%%%%%%%%%%%%%%%%%%%%%%%%%%%%%%%%%%%%%%%%%

\section*{Acknowledgments}

We thank the referee for his/her comments, which helped to improve
the manuscript. CT and LH acknowledge funding from the INAF
PRIN-SKA 2017 program 1.05.01.88.04. NRN acknowledges financial
support from the ``One hundred top talent program of Sun Yat-sen
University'' grant N. 71000-18841229. CS acknowledges financial
support from the Oxford Hintze Centre for Astrophysical Surveys.
LM acknowledges the support from the grants PRIN-MIUR 2017 WSCC32
and ASI n.2018-23-HH.0. DS is a member of the International Max
Planck Research School (IMPRS) for Astronomy and Astrophysics at
the Universities of Bonn and Cologne. MS acknowledges financial
support from the VST project (P.I. P. Schipani). GD acknowledges
support from CONICYT project Basal AFB-170002.
Based on data products from observations made with ESO Telescopes
at the La Silla and Paranal Observatory under programme IDs
177.A-3016, 177.A-3017, and 177.A-3018, as well as on data
products produced by Target/OmegaCEN, INAF-OACN, INAF-OAPD, and
the KiDS production team, on behalf of the KiDS consortium.
OmegaCEN and the KiDS production team acknowledge support by NOVA
and NWO-M grants.

\bibliographystyle{aa}
%\bibliography{D:/Documenti/latex/Bibtex/myrefs,D:/Documenti/latex/Bibtex/myrefs_KiDS_add,D:/Documenti/latex/Bibtex/myrefs_ESKAPEHI}

\end{document}